\documentclass[aps,pra,reprint]{revtex4-2}
\usepackage{colordvi,epsfig,color,amsmath,amssymb,bm}

\def\be{\begin{equation}}
\def\ee{\end{equation}}
\def\bee{\begin{eqnarray}}
\def\eee{\end{eqnarray}}

\def\kb{k_{\rm B}}
\def\tilde{\widetilde}

\def\halb{\mbox{$\frac{1}{2}$}}

\newcommand{\bbbone}{{\mathchoice {\rm 1\mskip -4mu l}{\rm 1\mskip 
-4mu l}{\rm 1\mskip -4.5mu l}{\rm 1\mskip -5mu l}}}

\begin{document}

\title{Dephasing and pseudo-coherent quantum dynamics in super-Ohmic environments}

\author{Ph. Nacke$^{1}$, F. Otterpohl$^{2}$, M. Thorwart$^{2,3}$, and P. Nalbach$^{1}$ 
}
\affiliation{
$^1$Fachbereich Wirtschaft \& Informationstechnik, Westf\"alische Hochschule, M\"unsterstrasse 265, 46397 Bocholt, Germany\\
$^2$I.\ Institut f\"ur Theoretische Physik, Universit\"at Hamburg, Notkestra{\ss}e 9, 22607 Hamburg, Germany\\
$^3$The Hamburg Centre for Ultrafast Imaging, Luruper Chaussee 149, 22761 Hamburg, Germany
}

\date{\today}

\begin{abstract}
Dephasing in quantum systems is typically the result of its interaction with environmental degrees of freedom. We investigate within a spin-boson model the influence of a super-Ohmic environment on the dynamics of a quantum two-state system. A super-Ohmic enviroment, thereby, models typical bulk phonons which are a common disturbance for solid state quantum systems as, for example, NV centers.  
By applying the numerically exact quasi-adiabatic path integral approach we show that for strong system-bath coupling, pseudo-coherent dynamics emerges, i.e., oscillatory dynamics at short times due to slaving of the quantum system to the bath dynamics. We extend the phase diagram known for sub-Ohmic and Ohmic environments into the super-Ohmic regime and observe a pronounced non-monotonous behaviour. Super-Ohmic purely dephasing fluctuations strongly suppress the amplitude of coherent dynamics at very short times with no subsequent further decay at later times. Nevertheless, they render the dynamics overdamped. The according phase separation line shows also a non-monotonous behaviour, very similar to the pseudo-coherent dynamics. 
\end{abstract}


\maketitle

\section{Introduction \& Motivation}

Dissipation, i.e., dephasing and relaxation, in a quantum system is a result of its coupling to environmental fluctuations.
At strong coupling the dissipative environment may also lead to fully incoherent dynamics or even complete suppression of the coherent quantum dynamics (localization).
Theoretical studies, typically, reduce the relevant quantum system to a paradigmatic two-state quantum system with the model Hamiltonian $H_0=\Delta \sigma_x/2$ with tunnel element $\Delta$ and Pauli matrices $\sigma_j$ (with $j=x,y,z$), interacting with harmonic degrees of freedom \cite{WeissBuch, Leggett1987} which act as the dissipative environment. 
The central characteristic of the environmental fluctuations is their spectral distribution which is typically modelled as a continuous function of frequency $\omega$, increasing $\propto \alpha \omega^s$ with spectral exponent $s$ and coupling strength $\alpha$. 
It is denoted as sub-Ohmic, Ohmic and super-Ohmic for $0<s<1$, $s=1$ and $s>1$ respectively.

Usually, one addresses {\it relaxational} fluctuations which cause transitions in the two-state system (via a coupling to $\sigma_z$) and thus relaxation. The according model is termed spin-boson model \cite{WeissBuch, Leggett1987} and shows in the Ohmic case with increasing $\alpha$ at $\alpha=\alpha_o(s=1)=\halb$ for the expectation value $P_z(t)=\langle\halb\sigma_z(t)\rangle$ (and likewise for the correlation function $\langle \sigma_z(t)\sigma_z(0)\rangle$) a dynamic transition from coherent oscillatory behavior to incoherent dynamics. 
Incoherent dynamics \cite{Leggett1987, WeissBuch} occurs when the oscillatory frequency is renormalized to zero at a finite system-bath coupling $\alpha$. The dynamics might be effectively overdamped (when the damping rate exceeds the oscillatory frequency) already at lower couplings. 
Increasing the coupling strength further to $\alpha=\alpha_c(s=1)=1$ the Ohmic spin-boson model exhibits at zero temperature a quantum phase transition into a localized phase with a degenerate ground state, i.e., the eigenstates to $\sigma_z$. 
Whereas a sub-Ohmic environment shows similar behaviour \cite{AndersSubOhm2007, WinterSubOhm2009, AlvermannSubOhm2009, NalbachSubOhmPRB2010, GuoSubOhm2012, AnkerholdSubOhmPRL2013, NalbachSubOhm2013} depending on the spectral exponent $s$, a super-Ohmic environment exhibits only damped oscillatory behavior \cite{WeissBuch, Leggett1987}. 

Super-Ohmic reservoirs receive fairly little consideration in theoretical studies since the dynamics turns neither localized nor incoherent even at strongest coupling except at high temperatures \cite{WuergerSuperOhm1997}. 
Super-Ohmic reservoirs are, however, fairly common and are, for example, the cause of damping for all dipolar defects in non-conducting solids, i.e., when phonons are the main noise source. 
Prominent examples are tunneling two-level systems in amorphous systems, glasses \cite{JaeckleGlaeser1972, NalbachJLTPTS2004} and crystals \cite{NalbachJoP2001} but also NV and SiV center in diamonds \cite{SuperOhmNVAlkauskas2014, SuperOhmNVJahnke2015, SuperOhmNVPRB2016}. Super-Ohmic reservoirs are also relevant noise sources for charge double quantum dots \cite{NalbachLZ2013} and for energy transfer in the FMO exciton transfer complex \cite{B777-Renger-Marcus-JCP-2002}. 

For the sub-Ohmic and the Ohmic bath, recently a particular dynamic behaviour at short times at strongest coupling was revealed (termed pseudo-coherent) exhibiting in $P_z(t)$ initially oscillatory behaviour with at least a single minimum \cite{Otterpohl2022}. 
This oscillatory polarization dynamics results from the two-state system being slaved to the bath which itself shows coherent dynamics on a time scale $\omega_c^{-1}$ due to a finite upper cut-off frequency $\omega_c$ of the environmental fluctuations. 
On any other time scale the quantum system shows no oscillatory (coherent) dynamics at these coupling strengths. 
In the scaling limit $\omega_c\to \infty$ the pseudo-coherent oscillatory behaviour shifts to earlier times and finally vanishes. 
Although relaxational super-Ohmic fluctuations \cite{Leggett1987, WeissBuch} cause neither localization nor overdamping, we show in this work numerically (as our first result) that for strong system-bath coupling pseudo-coherent dynamics, i.e., oscillatory dynamics at short times due to slaving the quantum system to bath dynamics, is present.  We determine the minimal coupling strength $\alpha_B(s)$, at which pseudo-coherent dynamics sets in, as function of the spectral exponent $s$ of the environmental fluctuations. This extends the phase diagram of the pseudo-coherent dynamics into the super-Ohmic regime. Surprisingly, $\alpha_B(s)$ as function of $s$ is nonmonotonic and shows a maximum at $s\simeq 2$.

In addition, we consider {\it purely dephasing} super-Ohmic fluctuations, i.e., the system-bath coupling operator is $\propto \sigma_x$. The according model is termed independent-boson model \cite{Mahan} and we find that purely dephasing super-Ohmic fluctuations do not cause exponential decay of coherence. Instead after an initial Gaussian decay \cite{BraunPRL2001} for a time scale $\omega_c^{-1}$ no further dephasing occurs at later times. The dynamics remains coherent but this strongly non-Markovian short-time behaviour severely diminishes its amplitude.
For strong system-bath coupling the dynamics is effectively overdamped and an effective transition coupling strength can be determined. Mapping out this transition as a function of $s$, it shows surprisingly a very similar nonmonotonic behaviour as the pseudo-coherent phase. Finally, we show that realistic quantum systems, which are exposed to both types of fluctuations, exhibit due to the purely dephasing fluctuations an initial fast dephasing. It subsequently saturates, and is followed by an additional exponentially decay due to the relaxational fluctuations. 

In the following, we present the studied model and shortly describe the used numerical methods in section II. In sections III, IV and V results for the pseudo-coherent behaviour, the Gaussian decay behavior and the mixed case are discussed before we conclude.

\section{Model \& numerical Method}

The Hamiltonian ($\hbar=1$)
\be\label{spin-boson} H \,=\, \frac{\Delta}{2}\sigma_x
\,+ (u_x\sigma_x + u_z\sigma_z )\sum_k\lambda_k\hat{q}_k 
\,+ H_B
\ee
describes a quantum two-state system with tunneling element $\Delta$ which is coupled by $\lambda_k$ to the displacements $\hat{q}_k$ of the harmonic environmental fluctuations $H_B=\halb\sum_k (\hat{p}_k^2+\omega_k^2\hat{q}_k^2 ) $ with frequency $\omega_k$. 
Herein, the case $u_x=0$ and $u_z=1$ reflects coupling to relaxational fluctuations (spin-boson model) whereas the case of purely dephasing fluctuations is given by $u_z=0$ and $u_x=1$ (independent-boson model). 

The spectral function of the fluctuations is 
\be
G(\omega) = \sum_k\frac{\lambda_k^2}{2\omega_k}\delta(\omega-\omega_k) 
= 2\alpha\omega_s^{1-s} \omega^s e^{-\omega/\omega_c} ,
\ee
with spectral exponent $s$ and a maximal environmental frequency (cut-off frequency) $\omega_c$. 
The frequency $\omega_s$ serves to keep the coupling strength $\alpha$ dimensionless and we fix $\omega_s=\omega_c$.

\begin{figure}[t!]
\includegraphics[width=80mm]{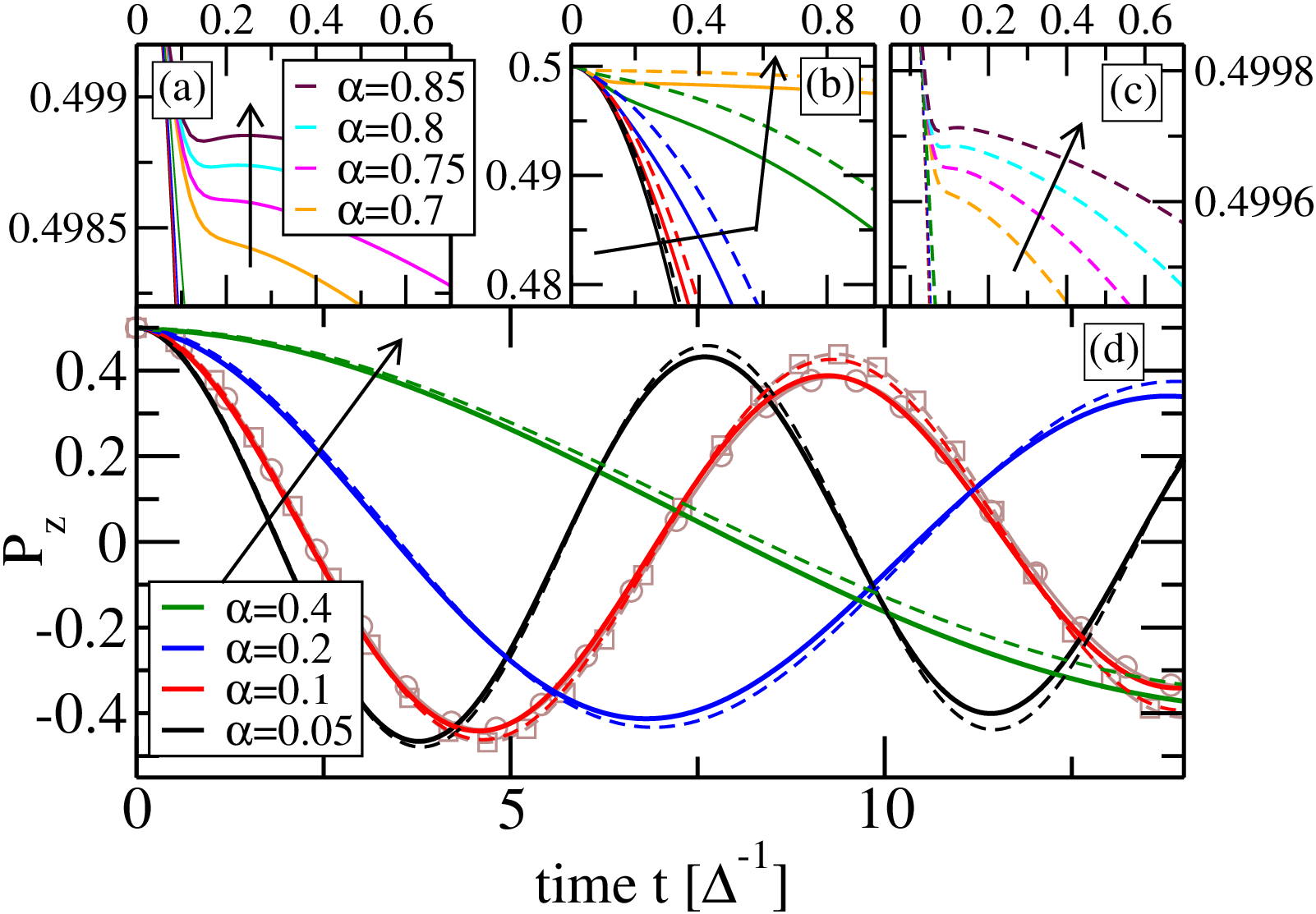}
\caption{\label{Bild1} (All graphs) Polarization $P_z(t)$ versus time for various values of the coupling strength $\alpha$ and a spectral exponent $s=2$ for $\omega_c=10\Delta$ (full lines) and $\omega_c=20\Delta$ (dashed lines) at $T=0$. The graphs show different time and amplitude regimes. The brown lines (full with circle symbols and dashed with square symbols) lines represent weak-coupling approximations \cite{WeissBuch} to the $\alpha=0.1$ results. The arrows indicate increasing coupling strength.}
\end{figure}

We calculate the time-dependent polarization $P_z(t)=\langle \halb\sigma_z\rangle_t$, using a factorizing initial preparation of the system with $P_z(0)=\halb $ and the thermal distribution of the bath at zero temperature. To determine $P_z(t)$ for relaxational fluctuations ($u_x=0$ and $u_z=1$) we use the numerically exact real-time quasiadiabatic propagator path integral (QUAPI) \cite{MakriJMP1995, MakriQUAPI1995a, MakriQUAPI1995b, PalmJCP2018}. Once the bath oscillators have been integrated out, an effective dynamics of the system arises which is nonlocal in time. To treat the highly entangled system-bath dynamics, we make use of the time-evolving matrix product operator (TEMPO) technique in terms of a numerically highly efficient tensor network \cite{StrathearnTNQuapi2018}.
Purely dephasing fluctuations are analytically tractable within the independent boson model \cite{Mahan} which was already successfully employed for the Ohmic case \cite{PalmaIndepBoson1996, JohnHenryIndepBoson2002, GoanIndepBoson2010} and the sub-Ohmic case \cite{NalbachSubOhm2013}. Accordingly, $P_z(t)$ is determined exactly in this case.

\section{Numerical results for the super-Ohmic spin-boson model}

In the following we study the influence of a relaxational super-Ohmic bath ($u_x=0$ and $u_z=1$) at zero temperature on the dynamics of a quantum two-state system.  
Fig. \ref{Bild1} shows the polarization $P_z(t)$ versus time for various values of the coupling strength $\alpha$ and a spectral exponent $s=2$ for $\omega_c=10\Delta$ (full lines) and $\omega_c=20\Delta$ (dashed lines). The polarization exhibits damped oscillations (see Fig. \ref{Bild1} (d)).
The oscillation frequency is renormalized due to the coupling to environmental fluctuations, i.e., with increasing coupling strength $\alpha$ the frequency is decreased \cite{WeissBuch}, following $\Delta_{\rm eff}=\Delta\exp(-\tilde{\alpha})$ with $\tilde{\alpha}=8\alpha\Gamma(s-1)$ with the Gamma function $\Gamma(x)$. Note that the frequency renormalization is independent of $\omega_c$. The observed weak damping is fully described by the one-phonon rate \cite{WeissBuch}  $\gamma_{\rm eff}=(\pi/2) \alpha \Delta_{\rm eff}^2 / \omega_c$ (for $s=2$ and $T=0$) which shows a dependence on $\omega_c$ (compare full with dashed lines in Fig. \ref{Bild1} (d)). The brown full and dashed lines (with circle and square symbols respectively) in Fig. \ref{Bild1} (d) are the analytical expections for the numerically determined red lines for $\alpha=0.1$ and we observe good agreement between both.
When focusing on small times at stronger couplings (see Fig. \ref{Bild1} (a), (b) and (c)) we observe at times roughly $\omega_c^{-1}$ the emergence of a minimum in the dynamics (focused on in Fig.\ref{Bild1} (a) for $\omega_c=10\Delta$ and (c) for $\omega_c=20\Delta$ showing coupling strengths $0.7$, $0.75$, $0.8$ and $0.85$).
This minimum is similar to the pseudo-coherent behaviour observed for $s\le 1$ in Ref. \cite{Otterpohl2022}. With increasing coupling, the pseudo-coherent minimum shifts towards earlier times. Since the dynamics is not localized for $s>1$, it is hard to resolve the shallow minimum and, thus, $\alpha_B(s)$. The accuracy depends on the maximally simulated time $t_{\rm max}$. For $1.0\le s\le 1.3$ we used $t_{\rm max}=3\Delta^{-1}$ and for larger $s$ we used $t_{\rm max}=\Delta^{-1}$.
For $\omega_c=20\Delta$ we observe qualitatively the same behaviour as for $\omega_c=10\Delta$ (compare Fig. \ref{Bild1} (a) and (c)). The absolute value of $P_z(t)$ at the pseudo-coherent minimum increases with $\omega_c$. Surprisingly, the transition coupling strength $\alpha_B$ does not change with $\omega_c$. 

\begin{figure}[t!]
\includegraphics[width=80mm]{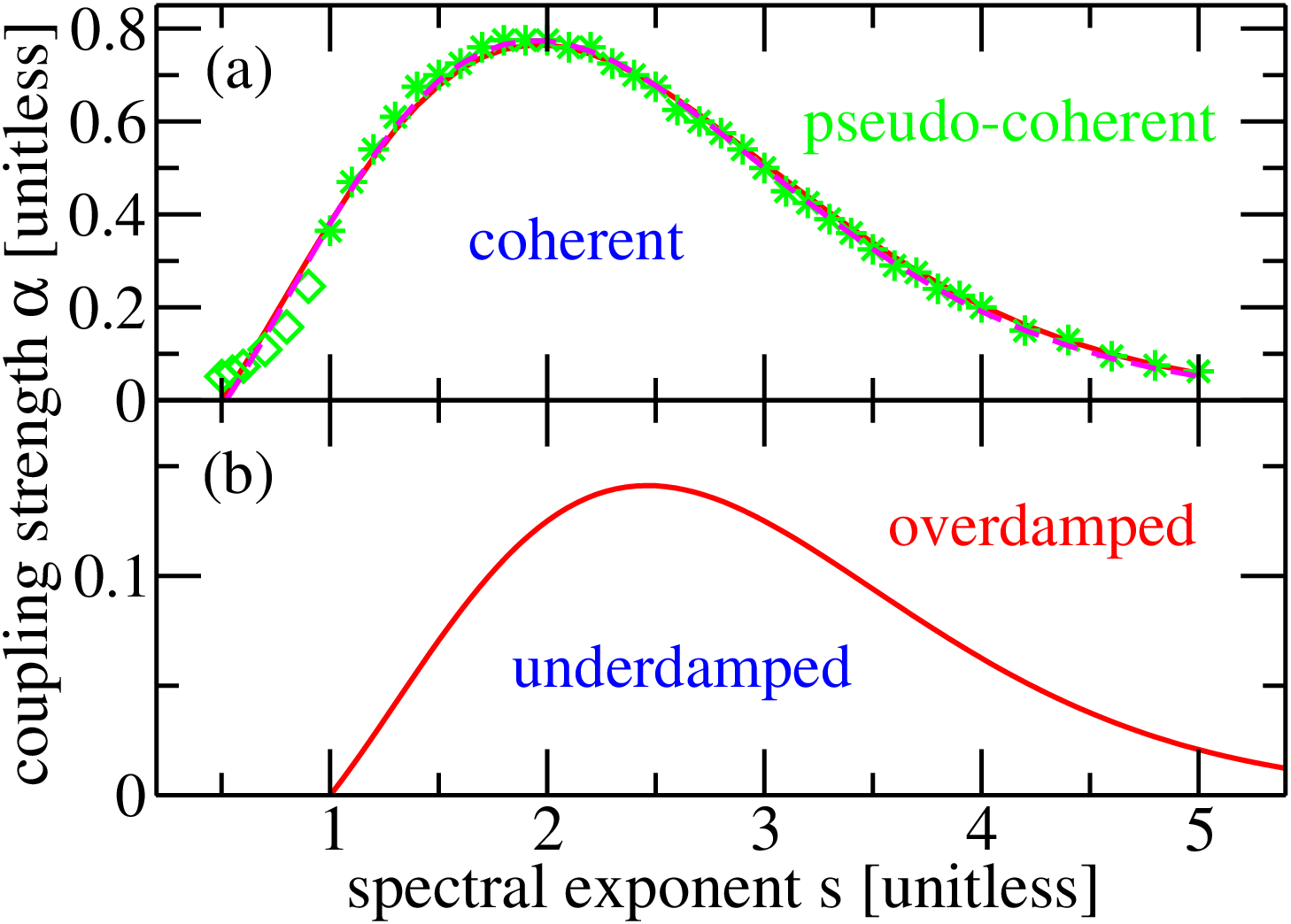}
\caption{\label{Bild2} (a) The green stars are the numerically determined transition coupling strength $\alpha_B(s)$ for the crossover to pseudo-coherent dynamics in the super-Ohmic relaxational bath. The green diamonds are the according data for $s<1$ taken from Ref. \cite{Otterpohl2022}. The red line is a fit with $A/(8\Gamma(s-\halb ))$ resulting in $A=5.42$ and the magenta line is a fit with $A/(8\Gamma(B\cdot s- C ))$ resulting in $A=5.487$, $B=1.026$ and $C=0.5342$. 
(b) The red full line reflects the transition $\alpha_o(s)$ to overdamped behaviour in the super-Ohmic purely dephasing bath. 
(Both graphs) Parameters are $\omega_c=10\Delta$ and $T=0$.}
\end{figure}

In Fig. \ref{Bild2} (a) we show the phase diagram of the super-Ohmic pseudo-coherent behaviour. The green stars are the numerically determined transition coupling strengths $\alpha_B(s)$ for the crossover to pseudo-coherent dynamics, i.e., where the minimum in the dynamics emerges. The green diamonds are the according data for $s<1$ taken from Ref. \cite{Otterpohl2022}. The phase separation line shows a strong non-monotonous behaviour with a peak for $s\simeq 2$. It smoothly connects to the results of Ref. \cite{Otterpohl2022}. Note the difference of a factor $\halb$ in the system-bath coupling in Hamiltonian (\ref{spin-boson}) which results in a factor $4$ difference between our coupling strength and the one given in Ref. \cite{Otterpohl2022}. The magenta line in Fig. \ref{Bild2} (a) is a fit with $A/(8\Gamma(B\cdot s- C ))$ resulting in optimal values $A=5.487$, $B=1.026$ and $C=0.5342$. The red line (which falls on top of the magenta line) is a fit with the simplified function $A/(8\Gamma(s-\halb ))$ resulting in $A=5.42$. Both fits reasonably describe the peak and the data for $s\gtrsim 1$, i.e., for the super-Ohmic regime. For $s\le 1$ (sub- and Ohmic regime) the fit does not describe the data sufficiently well.

\section{Analytical results for the super-Ohmic independent-boson model}

The case of a purely dephasing bath with diagonal coupling ($u_x=1$ and $u_z=0$) can be studied analytically.
Employing the transformation $T_P=e^{i\psi\sigma_x}$ with $\psi=\sum_k\frac{\lambda_k}{\omega_k^2}\hat{p}_k$ results in $H_P=T_P^\dagger HT_P=\Delta \sigma_{x,P}/2 + H_B$ with $\sigma_{x,P}=T^\dagger_P\sigma_x T_P = \sigma_x$.
Thus, the propagator $U_P=e^{-iH_Pt}$ is determined by a direct product of system and bath operators. At the same time, $\sigma_{z,P} = T^\dagger_P\sigma_zT_P = \sigma_z \cos\psi- \sigma_y\sin\psi$.
When measuring the polarization, we initially displace the system fully, i.e.,$P_z(0)=\halb$. Thus, $\langle \sigma_x\rangle =0$ and if the bath is allowed to equilibrate to this situation before the experiment starts, we can assume a factorized initial condition $\rho_{0,P_z} = \rho(t=0) = \halb (\bbbone + \sigma_z)\otimes \rho_{B,eq}$ with $\rho_{B,eq} = Z_B^{-1} e^{-\beta H_B}$ and $Z_B={\rm Tr}\{ e^{-\beta H_B} \}$ and $\beta=(\kb T)^{-1}$ for temperature $T$. A tedious calculation then results in the polarization
\be
P_z(t) = \halb \cos(\Delta t) e^{-\Gamma_T(t)},
\ee
with the decay function 
\be
\Gamma_T(t) = 4 \int_0^\infty d\omega\,  \frac{G(\omega)}{\omega^2}\left[ 1 - \cos\omega t \right] \coth(\beta\omega/2). 
\ee
Note that no frequency renormalization occurs here and that the decay function is dominated by high-frequency modes, in contrast to pure dephasing decay for sub-Ohmic and Ohmic baths. At zero temperature we find for super-Ohmic fluctuations $s>1$: 
\be 
\Gamma_0(t,s) = \tilde{\alpha}\,   \left\{ 1 - \frac{\cos\left[ (s-1) \arctan(\omega_c t) \right]}{\left[ 1+(\omega_ct)^2 \right]^{(s-1)/2} } \right\} , 
\ee
with the effective coupling $\tilde{\alpha}=8\alpha \Gamma(s-1)$
and the Gamma function $\Gamma(x)$. 

Initially, at times $\omega_ct\ll 1$ (and $(s-1)\lesssim 10$), we observe a spectral diffusion type of Gaussian decay: $\Gamma_0(t,s)\simeq \tilde{\alpha} \halb (s^2-s+2) (\omega_c t)^2$. At later times, $\omega_ct\gtrsim 1$, the decay function becomes constant, i.e., $\Gamma_0(t,s) \simeq \tilde{\alpha}$, and no further dephasing takes place. 
Thus, as long as $\tilde{\alpha}\ll 1$ dephasing is negligible. If $\tilde{\alpha}\gtrsim 1$, however, even though $\Gamma_0(t,s)$ becomes constant, dephasing supresses the response to negligible values. Thus, although strictly speaking the dynamics is oscillatory, the amplitude is vanishingly small and the dynamics is effectively overdamped, i.e., not a single sizable oszillation takes place. Defining $\tilde{\alpha}_o\equiv 1$ as the transition point, we find
\be \label{eq4}
\alpha_o(s) = \frac{1}{8\Gamma(s-1)} 
\ee
which we plot in Fig. \ref{Bild2} (b) by the full red line. We observe a strongly non-monotoneous behaviour with a peak at roughly $s \simeq 2.5$. Surprisingly, this peak strongly resembles the observed behaviour for the transition to pseudo-coherent dynamics although the peak is slightly shifted and the maximal value is considerably smaller. Testing this observation we employed the fit function $A/(8\Gamma(B\cdot s- C ))$ with fit parameters $A$, $B$ and $C$ above to fit $\alpha_B(s)$ as function of $s$ resulting in the magenta line in Fig. \ref{Bild2} (a). Fixing $B=1$ and $C=0.5$ does not deteriorate the fit and then we obtain $A=5.42$ (red line in Fig. \ref{Bild2} (a)). Note that $\alpha_o(s)$ neither depends on $\omega_c$, similar to our observation for $\alpha_B(s)$.

%
%

%
\begin{figure}[t!]
\includegraphics[width=80mm]{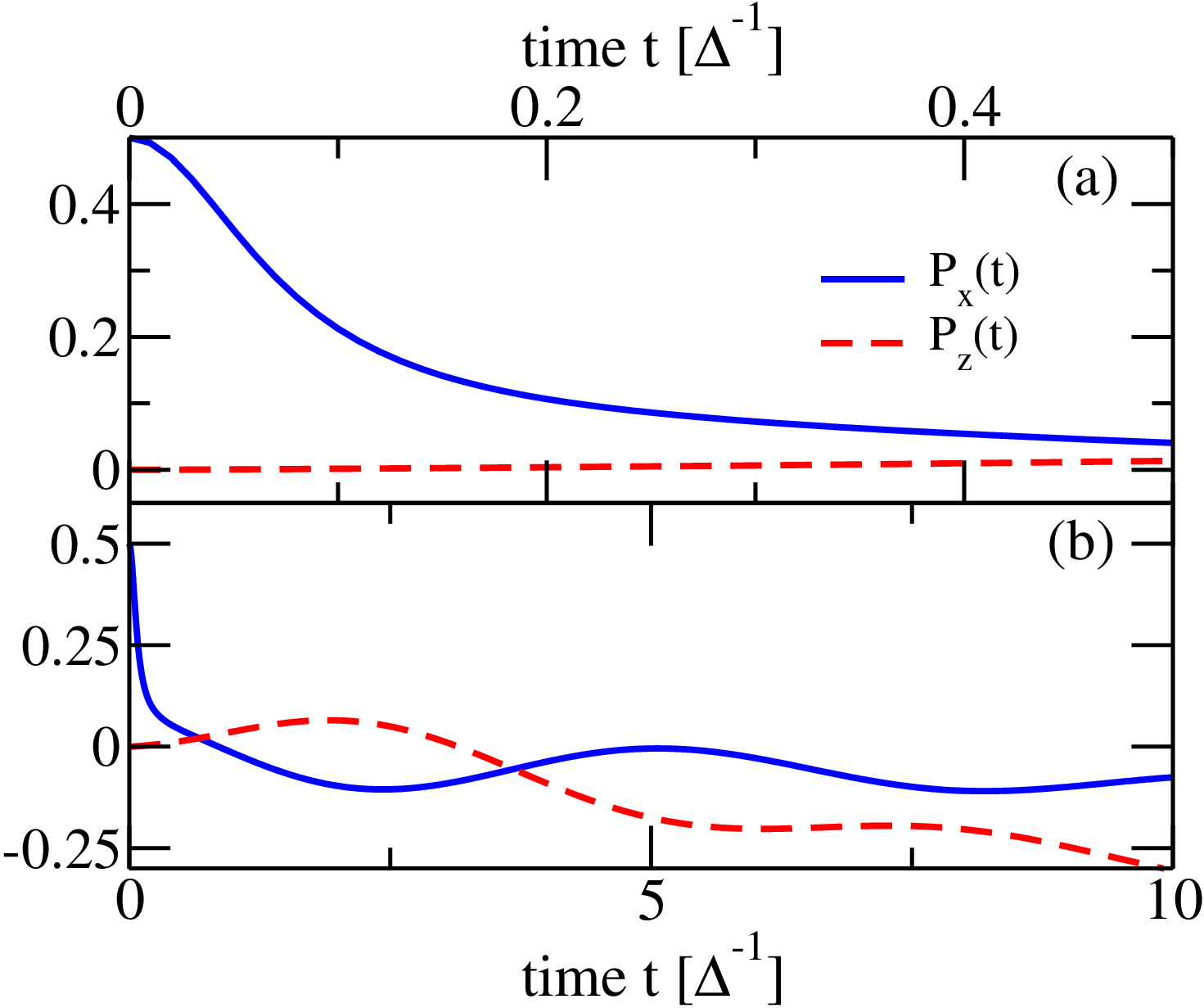}
\caption{\label{Bild3} Polarization $P_z(t)$ and $P_x(t)$ in the asymmetric super-Ohmic spin-boson model versus time for coupling strength $\alpha=0.2$, a spectral exponent $s=2$ and $\omega_c=10\Delta$ at $T=0$. (a) $P_z(t)$ and $P_x(t)$ are displayed at short times $t\le 0.5\Delta^{-1}$. (b) $P_z(t)$ and $P_x(t)$ are displayed for times $t\le 10\Delta^{-1}$.}
\end{figure}

\section{Experimental relevance of purely dephasing fluctuations}

The influence of relaxational super-Ohmic fluctuations on the dynamics of a quantum systems is readily obserservable in the population decay as well as in the form of decoherence. Dephasing due to super-Ohmic purely dephasing fluctuations is restricted to times much shorter than typical system times as for most solid state systems $\omega_c^{-1}\ll\Delta^{-1}$ holds. This dephasing is accordingly irrelevant for the system dynamics and for typical ensemble measurements it is simply countered by enlarging the ensemble. Thus, often theoretical studies, focussed on weak Markovian environments, neglect super-Ohmic purely dephasing completely.
For qubit applications, however, survival of coherence of single systems, for example, in NV centers, after initial preparation is key. Dephasing at any time scale shorter than the {\it calculation time} of the qubit, typically much longer than $\Delta^{-1}$, deteriorates qubit applications. 

It remains to study whether a system under the influence of both, relaxational and purely dephasing fluctuations, which seems experimentally the most likely situation, suffers the dephasing behavior of purely dephasing fluctuations as discussed above. The Hamiltonian in Eq. (\ref{spin-boson}) with $u_x=\sin\phi \not= 0 \not= u_z=\cos\phi $ can be transformed to
\[ H \,=\, \frac{\Delta}{2}\cos\phi \; \sigma_x + \frac{\Delta}{2}\sin\phi \; \sigma_z \,+ \sigma_z \sum_k\lambda_k\hat{q}_k 
\,+ H_B,
\]
i.e., an asymmetric spin-boson model. Fig. \ref{Bild3} plots the polarizations $P_z(t)=\langle\halb\sigma_z\rangle_t$ and $P_x(t)=\langle\halb\sigma_x\rangle_t$ in the transformed basis versus time for coupling strength $\alpha=0.2$, a spectral exponent $s=2$, $\omega_c=10\Delta$ and $\Delta\cos\phi=1=\Delta\sin\phi $.
In this transformed basis both, $P_z(t)$ and $P_x(t)$, exhibit thermalization and decoherence features. $P_x(t)$ shows at short times (Fig. \ref{Bild3} (a)), i.e., $t\lesssim 0.1\Delta^{-1} = \omega_c^{-1}$, a sharp Gaussian type decay down to roughly $e^{-\tilde{\alpha}}$. At later times (Fig. \ref{Bild3} (b)), the oscillation amplitude decays further but with a much smaller rate which is comparable to the dephasing rate observed in the symmetric spin-boson model for similar parameters (see Fig. \ref{Bild1}). Thus, indeed the dynamics in the asymmetric spin-boson model exhibits dephasing features of both, the relaxational and the purely dephasing fluctuations independently observed above.

\section{Conclusions}

We have investigated the polarization dynamics of a quantum two-state system coupled to a super-Ohmic environment within a spin-boson model. 
Super-Ohmic environments are fairly common in solid state quantum systems as they model typical bulk phonons. A prominent quantum system subject to super-Ohmic environments are NV centers \cite{SuperOhmNVAlkauskas2014, SuperOhmNVJahnke2015, SuperOhmNVPRB2016}.  
For the treatment of relaxational environmental fluctuations and the study of the polarization dynamics we employ the numerical exact quasi-adiabatic path integral approach combined with an efficient tensor network treatment. Purely dephasing fluctuations alone are treated analytically. The combination of both requires again numerical treatment.

Super-Ohmic environments can neither turn the dynamics localized nor incoherent. On the time scale of the bare quantum system, however, the dynamics is severely slowed down since the oscillation frequency of the polarization is strongly decreased with increasing system-bath coupling. At strong coupling we observe pseudo-coherent dynamics, i.e., oscillatory dynamics at short times due to slaving the quantum system to bath dynamics. We map the minimal coupling strength $\alpha_B(s)$, at which pseudo-coherent dynamics occurs, as function of the spectral exponent $s$ of the environmental fluctuations and, thus, extend the phase diagram of the pseudo-coherent dynamics \cite{Otterpohl2022} into the super-Ohmic regime. Surprisingly, $\alpha_B(s)$ as function of $s$ is nonmonotonic with a maximum at $s\simeq 2$.
Purely dephasing super-Ohmic fluctuations cause an initial Gaussian decay for a time scale $\omega_c^{-1}$ and then no further dephasing. Nevertheless the coherence amplitude is severely diminished at strong coupling rendering the dynamics effectively overdamped. The according phase separation line of coupling versus spectral exponent exhibits also a nonmonotonous behaviour, very similiar to the pseudo-coherent phase.

Realistic quantum systems are typically exposed to both types of environmental fluctuations, i.e., purely dephasing and relaxational ones. We show that the polarization dynamics then exhibits a fast initial Gaussian decay followed by the much slower (for the same system-bath coupling) exponential decay due to the relaxational fluctuations. Hence, neglecting the non-Markovian Gaussian decay due to the purely dephasing fluctuations may not be justified when studying the coherence of a quantum system as is relevant for quantum devices. Even at weak system-bath coupling $\alpha$ the inflicted decay is proportional $\propto \exp(-\tilde{\alpha})$ with $\tilde{\alpha}=8\alpha \Gamma(s-1)$ and the Gamma function $\Gamma(x)$ and might well be relevant for the envisioned quantum device. Since this super-Ohmic pure depahsing decay is dominated by high frequency environmental modes, dynamical decoupling schemes are also less effective as against, for example, $1/f$ noise.

\section*{Acknowledgement}

M.T. acknowledges support by the Cluster of Excellence CUI: Advanced Imaging of Matter of the Deutsche Forschungsgemeinschaft (DFG) – EXC 2056 – project ID 390715994.


\end{document}